\documentclass[twocolumn,shortnote]{jpsj3}
\usepackage{txfonts}
\usepackage{color}
\usepackage{ulem}

\title{
Enhanced Superconductivity up to 43 K by P/Sb Doping of Ca$_{1-x}$La$_x$FeAs$_2$}

\author{
\name{Kazutaka Kudo}$^{1,2}$\thanks{E-mail: kudo@science.okayama-u.ac.jp}, 
\name{Tasuku Mizukami}$^{1}$, 
\name{Yutaka Kitahama}$^{1}$, 
\name{Daisuke Mitsuoka}$^{1}$, 
\name{Keita Iba}$^{1}$, 
\name{Kazunori Fujimura}$^{1}$, 
\name{Naoki Nishimoto}$^{1}$, 
\name{Yuji Hiraoka}$^{1,2}$, and 
\name{Minoru Nohara}$^{1,2}$
}

\inst{
$^1$Department of Physics, Okayama University, Okayama 700-8530, Japan\\
$^2$Research Center of New Functional Materials for Energy Production, Storage and Transport, Okayama University, Okayama 700-8530, Japan
}

\begin{document}
\maketitle

A number of iron-based superconductors have been discovered,\cite{Ishida,Paglione,Johnston} which include 
LaFeAsO (1111-type structure),\cite{Kamihara} BaFe$_2$As$_2$ (122-type),\cite{Rotter}  LiFeAs (111-type),\cite{Tapp} and FeSe (11-type),\cite{Hsu}
as well as compounds with complex oxide spacer layers \cite{Kawaguchi,Zhu2,Ogino,Shirage} 
and arsenide spacer layers such as Ca$_{10}$(Pt$_4$As$_8$)(Fe$_2$As$_2$)$_5$. \cite{Kakiya,Ni1,Lohnert,Nohara,Heike,Kudo1}
The maximum superconducting transition temperature $T_{\rm c}$ is 55 K of the 1111-type structure. \cite{Ren}
In order to further increase $T_{\rm c}$, an exploration of novel structure types should be performed.

Very recently, Katayama {\it et al.} \cite{Katayama} and Yakita {\it et al.} \cite{Yakita} have reported superconductivity in Ca$_{1-x}$La$_x$FeAs$_2$ and Ca$_{1-x}$Pr$_x$FeAs$_2$, respectively, with a novel 112-type structure.
Ca$_{1-x}$La$_x$FeAs$_2$ crystalizes in a monoclinic structure with the space group $P2_1$ (No. 4) and consists of alternately stacked Fe$_2$As$_2$ and arsenic zigzag bond layers. \cite{Katayama}
Although pure CaFeAs$_2$ was not obtained, Katayama {\it et al.} found that the substitution of a small amount of La for Ca stabilizes the 112 phase 
and induces superconductivity at $T_{\rm c}$ = 34 K for $x$ = 0.16. 
Interestingly, Katayama {\it et al.}\cite{Katayama} suggested that the trace superconductivity of Ca$_{1-x}$La$_x$FeAs$_2$ could exhibit $T_{\rm c} =$ 45 K.

In this paper, we report that a large increase in $T_{\rm c}$ occurs with the phosphorus or antimony doping of Ca$_{1-x}$La$_x$FeAs$_2$. 
P-doped Ca$_{0.84}$La$_{0.16}$FeAs$_2$ and Sb-doped Ca$_{0.85}$La$_{0.15}$FeAs$_2$ exhibited $T_{\rm c}$ values of 41 and 43 K, respectively, while P/Sb-free Ca$_{0.85}$La$_{0.15}$FeAs$_2$ exhibited $T_{\rm c} = $ 35 K.

\begin{figure}[t]
\begin{center}
\includegraphics[width=8.5cm]{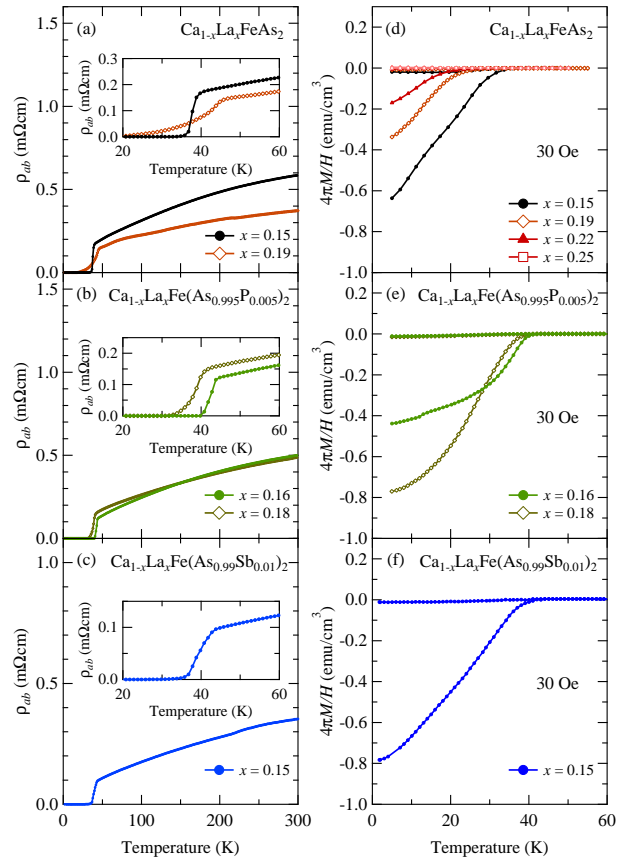}
\caption{\label{fig1} 
(Color online) Temperature dependences of the electrical resistivity parallel to the $ab$-plane, $\rho_{ab}$, and the magnetization $M$ measured at a magnetic field $H$ of 30 Oe for Ca$_{1-x}$La$_{x}$Fe(As$_{1-y}Pn_y$)$_2$ ($Pn =$ P and Sb). 
The insets show a magnified view in the vicinity of the superconducting transition.}
\end{center}
\end{figure}

Single crystals of Ca$_{1-x}$La$_{x}$Fe(As$_{1-y}Pn_y$)$_2$ ($Pn =$ P and Sb) were grown by heating a mixture of Ca, La, FeAs, As, P, and Sb powders. 
A stoichiometric amount of the mixture was placed in an aluminum crucible and sealed in an evacuated quartz tube. 
The preparation was carried out in a glove box filled with argon gas. 
Ampules were heated at 700 $^\circ$C for 3 h, heated to 1100 $^\circ$C at a rate of 46 $^\circ$C/h, and cooled to 1050 $^\circ$C at a rate of 1.25 $^\circ$C/h, followed by furnace cooling. 
The obtained samples were characterized by powder X-ray diffraction (XRD) analysis, performed using a Rigaku RINT-TTR III X-ray diffractometer with Cu$K_{\alpha}$ radiation. 
The Ca$_{1-x}$La$_{x}$Fe(As$_{1-y}Pn_y$)$_2$ was obtained together with a powder mixture of LaAs, FeAs, FeAs$_2$, and CaFe$_2$As$_2$. 
We separated platelike single crystals of the present system with typical dimensions of $0.4 \times 0.4 \times 0.02$ mm$^3$ from the mixture. 
The La content $x$ was analyzed by energy-dispersive X-ray spectrometry (EDS). 
The synthesis for nominal $x =$ 0.07--0.50 yielded samples with $x =$ 0.15--0.25. 
On the other hand, the P and Sb contents could not be determined because the nominal amounts of P ($y =$ 0.005) and Sb ($y =$ 0.01) were very small. In the rest of this paper, we assume the nominal values of $y$. 
Electrical resistivity $\rho_{ab}$ (parallel to the $ab$-plane) measurements were carried out by a standard DC four-terminal method in a Quantum Design PPMS. The magnetization $M$ was measured using a Quantum Design MPMS.

We demonstrate the bulk superconductivity at 35 K and the trace superconductivity at 45 K in Ca$_{1-x}$La$_{x}$FeAs$_2$ using the temperature dependences of the electrical resistivity $\rho_{ab}$ and magnetization $M$ shown in Figs. 1(a) and 1(d), respectively. 
The electrical resistivity $\rho_{ab}$ of Ca$_{1-x}$La$_{x}$FeAs$_2$ with $x =$ 0.15 exhibits a sharp drop below 40 K, and zero resistivity is observed at 35 K. 
The diamagnetic behavior below $T_{\rm c} =$ 35 K clearly supports the emergence of bulk superconductivity. 
On the other hand, the $\rho_{ab}$ of Ca$_{1-x}$La$_{x}$FeAs$_2$ with $x = 0.19$ exhibits the onset of superconductivity at 45 K, as shown in Fig. 1(a), but no diamagnetic signal is observed at $\sim$45 K, as shown in Fig. 1(d).

We found that a small amount of isovalent doping converts this trace superconductivity into bulk superconductivity. 
As shown in Fig. 1(b), the P-doped Ca$_{1-x}$La$_{x}$Fe(As$_{0.995}$P$_{0.005}$)$_2$ shows the onset of superconductivity at 44 K and zero resistivity at 40 K for $x =$ 0.16. 
Evidence for bulk superconductivity is found in the magnetization $M$, which clearly shows diamagnetic behavior below $T_{\rm c} =$ 41 K, as shown in Fig. 1(e). 
The shielding volume fraction at 5 K corresponds to 44\% for perfect diamagnetism, supporting the emergence of bulk superconductivity. 
In a composition with $x =$ 0.18, although the $T_{\rm c}$ of 39 K is somewhat lower than that with $x =$ 0.16, the shielding volume fraction exhibits a high value of 77\%. 
The Sb-doped Ca$_{1-x}$La$_{x}$Fe(As$_{0.99}$Sb$_{0.01}$)$_2$ also exhibits the onset of superconductivity and zero resistivity at high temperatures such as 44 and 35 K, respectively, for $x =$ 0.15 [Fig. 1(c)]. 
In this case, the bulk superconductivity is further supported by diamagnetic behavior below $T_{\rm c} =$ 43 K, as shown in Fig. 1(f). 
The shielding volume fraction at 1.8 K corresponds to 78\% for perfect diamagnetism.

\begin{figure}[t]
\begin{center}
\includegraphics[width=7.9cm]{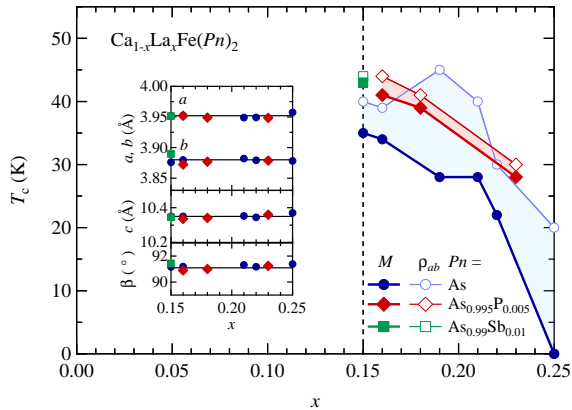}
\caption{\label{fig2} 
(Color online) $x$ dependence of the superconducting transition temperature $T_{\rm c}$, below which the magnetization shows a diamagnetic signal and the resistivity starts to drop in Ca$_{1-x}$La$_{x}$Fe(As$_{1-y}Pn_y$)$_2$ ($Pn =$ P and Sb). 
The inset shows the $x$ dependences of the lattice parameters.
}
\end{center}
\end{figure}

Figure 2 shows the $x$ dependences of $T_{\rm c}$ derived from the magnetization and the onset $T_{\rm c}$ derived from the electrical resistivity. 
The resistive-onset $T_{\rm c}$ determines the upper limit of the superconducting transition. 
In Ca$_{1-x}$La$_x$FeAs$_2$ not subjected to P/Sb doping, the $T_{\rm c}$ determined from the magnetization is considerably lower than the resistive-onset $T_{\rm c}$. 
On the other hand, in the case of additional P or Sb doping, these $T_{\rm c}$ values almost coincide with each other. 
Moreover, the magnetic $T_{\rm c}$ almost reaches the maximum resistive $T_{\rm c}$ (45 K) observed in Ca$_{1-x}$La$_x$Fe$_2$As$_2$ ($x =$ 0.19), suggesting that the small amount of isovalent doping has induced the potential high $T_{\rm c}$ phase in this $x$ range.

The enhancement of bulk superconductivity by isovalent P doping has been reported for the 122-type iron-based superconductor.\cite{Kudo2} 
The electrical resistivity of the La-doped Ca$_{1-x}$La$_x$Fe$_2$As$_2$ exhibited a superconducting transition at high temperatures such as $\sim$40--43 K, but no visible diamagnetic signal was observed.\cite{Saha1,Gao} 
In contrast, additional P doping induced bulk superconductivity at 45 K in Ca$_{1-x}$La$_x$Fe$_2$(As$_{1-y}$P$_y$)$_2$.\cite{Kudo2} 
In the case of La-and-P-codoped CaFe$_2$As$_2$, it has been suggested that independent tunings of the concentration of charge carriers by La doping and the cell volume by P doping optimize the superconductivity. 
Detailed studies of the crystal structure, which include the crystallographic sites of doped P/Sb, will provide useful information on the factors causing the enhancement of $T_{\rm c}$ in the present 112 system.

In summary, we report bulk superconductivity at 43 K induced by the isovalent doping of Ca$_{1-x}$La$_x$FeAs$_2$. 
We found that a small amount of isovalent P or Sb doping markedly enhances $T_{\rm c}$: P-doped Ca$_{0.84}$La$_{0.16}$FeAs$_2$ and Sb-doped Ca$_{0.85}$La$_{0.15}$FeAs$_2$ exhibit $T_{\rm c}$ values of 41 and 43 K, respectively. 
Note that if we can reduce $x$ to below 0.15, $T_{\rm c}$ will increase because the $T_{\rm c}$ of Ca$_{1-x}$La$_x$FeAs$_2$ tends to increase with decreasing $x$.
We should develop chemical methods for doing so.

\begin{acknowledgments}
We are grateful to N. Katayama and H. Sawa for valuable discussions. 
Part of this work was performed at the Advanced Science Research Center at Okayama University.
This work was partially supported by a Grant-in-Aid for Scientific Research (C) (25400372) from Japan Society for the Promotion of Science (JSPS), the Funding Program for World-Leading Innovative R\&D on Science and Technology (FIRST Program) from JSPS, 
and the program for promoting the enhancement of research universities from MEXT, Japan.
\end{acknowledgments}

\end{document}